\documentclass{svjour2}

\usepackage{units}
\usepackage{amsmath}
\usepackage{amssymb}

\newcommand{\weak}[1]{\ensuremath{\frac{\left\langle x\right| #1 \left|\psi\right\rangle }{\left\langle x|\psi\right\rangle }}}
\newcommand{\inlineweak}[1]{{\ensuremath{_{x}#1_{\psi}}}}
\newcommand{\weakvar}{\mathbb{V}_{p}\left(p|x\right)}

\journalname{Foundations of Physics}

\title{How the Weak Variance of Momentum can turn out to be Negative}

\author{M. R. Feyereisen}
\institute{M. R. Feyereisen \at
	     University of Amsterdam\\
	      Science Park 904, 1098 XH Amsterdam, Netherlands\\
               \email{m.r.feyereisen@uva.nl}
}
\date{}

\begin{document}

\maketitle

\begin{abstract}
Weak values are average quantities, therefore investigating their associated variance is crucial in understanding their place in
quantum mechanics. We develop the concept of a position-postselected \emph{weak variance} of momentum as cohesively as possible, building primarily on material from  \cite{moyal1949quantum,WignerHydro}.

The weak variance is defined in terms of the Wigner function, using a standard construction from probability theory. We show this corresponds to a measurable quantity, which is not itself a weak value. It also leads naturally to a connection between the imaginary part of the weak value of momentum and the quantum potential. We study how the negativity of the Wigner function causes negative weak variances, and the implications this has on a class of `subquantum' theories. We also discuss the role of weak variances in studying determinism, deriving the Classical Limit from a variational principle.


\end{abstract}

\newpage

\section{Weak Values and Variances}
\subsection{Weak Values}
\subsubsection{Introduction}
In the two state vector formalism, a weak value is equal to the transition amplitude between a prepared initial state and a postselection state \cite{aharonov2002two}. We focus in this study on the weak value
and the ``weak variance'' of the momentum operator $\hat{p}$
postselected by position $\left\langle x\right|$,
\[
\inlineweak{\hat{p}} =\weak{\hat{p} }\; ,\quad \inlineweak{\hat{p}} \in\mathbb{C}.
\]
A number of the following results should generalise to any
pair of canonically noncommuting variables, although a general definition of the weak variance is not the objective of this study.
The weak value $\inlineweak{\hat{p}} \in \mathbb{C}$ can be split
into real and imaginary parts \cite{hiley2012weak} as
\begin{equation}
\inlineweak{\hat{p}} = \weak{\hat{p}}=\hbar \partial_x \left( \arg\psi\right)-i\left(\frac{\hbar}{2}\frac{\partial_{x}\left|\psi\right|^{2}}{\left|\psi\right|^{2}}\right). \label{eq:weakvaluep}
\end{equation}

One might naively attempt to define a weak variance as the second central moment out of weak values $\inlineweak{\hat{p}^2} - \inlineweak{\hat{p}}^2$. However, the fact that weak values are complex requires one to take real parts of this naive construction -- and it makes a difference whether one squares the second term before or after taking real parts: $\mathrm{Re}(\inlineweak{\hat{p}}^2) \neq (\mathrm{Re}\,\inlineweak{\hat{p}})^2$. This ambiguity motivates the search for a more `fundamental' approach to these statistical quantities, which this first section will attempt to describe.

This study is organised as follows. We begin by reviewing the relation between the real parts of weak values and the phase space formalism of quantum mechanics. The quasistatistical interpretation of this formalism suggests a natural and unique definition for the ``weak variance'' in terms of the Wigner function. In \textsection 2, the embedding of configuration space formalisms into the phase space formalism then allows the weak variance to be related to the thermodynamics of the Madelung fluid, to the de Broglie--Bohm quantum potential $Q$, and to an experimentally measurable combination $\mathrm{Re}(\inlineweak{\hat{p}^2}-\inlineweak{\hat{p}}^2)$ of weak values.

In \textsection 3 we discuss the titular problem, that the negativity of the Wigner function can result in a negative weak variance. The weak variance therefore does not define a standard deviation of weak measurements from the weak value. The sign of the weak variance is shown to be related to the local extrema of the probability amplitude. The relation between weak variances and determinism paves the way for a detailed study of the Classical Limit in \textsection 4. We derive the classical limit from a variational principle involving the average stochastic contribution to the total momentum variance, and demonstrate the existence of a semiclassical limit where particle motion occurs, on average, classically.

\subsubsection{The importance of the measurement apparatus}

It is important to note \cite{WeakParrott} that weak values are not measured independently of an interaction Hamiltonian or a meter system. The real part of a weak value is well known to be the conditional average one would measure in the ideal limit of zero disturbance form the measuring apparatus \cite{dressel2012significance,dressel2010contextual,dressel2012contextual}. The imaginary part of a weak measurement is unrelated to observables, it arises from the disturbance due to coupling with a measurement apparatus \cite{dressel2012significance}. As such, it is not uniquely defined and not intrinsic to the measured system \cite{WeakParrott}.

Since only the real part informs us about our operator as an observable, we opt to describe the variance of only the real part of the weak value in this study. However, the imaginary part in \eqref{eq:weakvaluep} will emerge naturally during our investigation of weak variances for the real part, without reference to a specific interaction Hamiltonian or meter system. 
In principle the `ideal' decomposition \eqref{eq:weakvaluep} would be measured as \cite{WeakParrott,Jozsa}
\begin{equation}
\inlineweak{\hat{p}} = \mathrm{Re}\left(\weak{\hat{p}}\right) + \eta \; \mathrm{Im}\left(\weak{\hat{p}}\right),
\label{eq:experimental_eta}
\end{equation} such that the theoretically natural imaginary part in \eqref{eq:weakvaluep} is arbitrarily scaled in experiments by adjusting the value of $\eta$. This parameter depends on meter states and observables, as well as the interaction strength with the apparatus \cite{WeakParrott}.

We shall therefore discuss the imaginary part \emph{as if} it were intrinsic to the system, absorbing all dependence on the measurement protocol into the parameter $\eta$. This has the disadvantage of illusorily disconnecting these results from the experimental setups that give rise to them, which comes hand in hand with the hidden advantage of being a more universal treatment than that derived from any specific experimental setup. As such, the weak variance we define is related to the stochastic nature of quantum theory; it is not the experimental variance of any apparatus (although it may manifest itself as such in a measurement setting). Not without irony, this apparatus-agnostic treatment will suggest multiple possible experimental setups to study the weak variance (\textsection \ref{manymeasurements}).

\subsubsection{Weak Values as Conditional Expectations}

In the quasistatistical interpretation \cite{moyal1949quantum,takabayasi1954formulation,muckenheim1986review}
of the phase space formalism \cite{wigner1932,groenewold,PhaseSpaceOverview,KimNoz}, the canonical observables $x,p\in\mathbb{R}$ are treated as random variables. Since these observable are incompatible, no joint probability distribution can be reasonably be given for them -- the phase space distributions are understood to be bivariate \emph{pseudoprobability} distributions. For example, the Wigner function $\mathcal{W}\left(x,p\right)$ is not positive definite. For a review of extended probability we refer the reader to \cite{muckenheim1986review}.

Our choice of the Wigner function to define the weak variance is motivated by this quasidistribution's special status amongst physically equivalent representations of the phase space formalism (which correspond to different choices of operator ordering, $[\hat{x},\hat{p}]\neq0$). Indeed, only the Wigner function produces the correct marginals and observable averages by direct integration \cite{PhaseSpaceOverview}; other quasidistributions require explicit convolutions over noncommutativity terms. As such, the Wigner function is uniquely singled out by the quasistatistical outlook, where such noncommutativity terms would interfere with the interpretation of observable averages as moments of the pseudodistribution.

The real part of a weak value is a conditional average \cite{dressel2010contextual,dressel2012contextual,wiseman2007grounding}. Hence, to obtain weak values of momentum, one must consider not the momentum $p$, but instead the mean value of $p$ at a given position, for a statistical ensemble. One may consequently express the experimental protocol of postselection with conditional random variables: the quantity $p|x$ ($p$ conditioned by $x$) is also a random variable. To find the distribution of $p|x$, we need to consider the conditional distribution $\mathcal{W}\left(p|x\right)$.
The Wigner function has a marginal distribution 
\begin{equation}
\mathcal{W}\left(x\right)=\int\mathcal{W}\left(x,p\right)dp=\left|\psi(x)\right|^{2},
\label{eq:marginalWigner}
\end{equation}
 and so the conditional distribution for the random variable $p|x$ is simply
\begin{equation}
\mathcal{W}\left(p|x\right)=\frac{\mathcal{W}\left(x,p\right)}{\mathcal{W}\left(x\right)}.
\label{eq:conditionalWigner}
\end{equation}

These observables $p$, $x$ are incompatible: in what sense can we condition one on the other if they are not meaningful simultaneously? A random variable is \emph{defined} by its distribution. Individual realisations of $x$ and $p$ are meaningful positions and momenta only because their distributions (the marginals of the Wigner quasidistribution) are proper distributions. The random variable $p|x$ is defined by the conditional quasidistribution $\mathcal{W}\left(p|x\right)$, which from \eqref{eq:conditionalWigner} is clearly not nonnegative. Hence individual realisations of $p|x$ may occur with negative probability. This follows the same logic as stating that joint realisations $(x,p)$ may occur with negative probability as realisations of the full Wigner function.

However, the fact that individual realisations of $p|x$ can occur with negative probabilities does not preclude the use of this conditional variable as an intermediate step in a calculation of a physical quantity \cite{muckenheim1986review}. Furthermore, the mean or the variance of $p|x$ need not be confined to intermediate steps in this way, since they are not random variables but instead contribute to the description of the physical state of an ensemble (i.e. they descibe the Wigner function as a whole rather than the outcome of a measurement).

The mean value of $p|x$, 
\begin{equation}
\mathbb{E}_{p}\left(p|x\right)\equiv\tilde{p},
\label{eq:RealPartDefinition}
\end{equation}
can be thought of as a function of $x$ \cite{takabayasi1954formulation}. This mean value $\tilde{p}$ will in \textsection \ref{derivation} be shown to equal the real part of the weak value $_{x}\hat{p}_{\psi}$. One may then express the conditional mean momentum in terms of $\mathcal{W}$
as the first conditional moment \cite{moyal1949quantum}, i.e. the first moment of the conditional distribution: 
\begin{eqnarray}
\tilde{p} & = & \int p\mathcal{W}\left(p|x\right)\; dp\label{eq:WignerDefinition}\\
 & = & \frac{1}{\mathcal{W}\left(x\right)}\int\; p\mathcal{W}\left(x,p\right)\; dp.
\label{eq:ExpandedWignerDefinition}
\end{eqnarray}

It is possible to formulate complex weak values of arbitrary observables directly in the phase space formalism using cross-Wigner functions \cite{deGosson2012weak}, but as noted earlier our focus is currently limited to the variance of the real part of position-postselected weak momentum measurements.

\subsection{Weak variance as the Second Conditional Cumulant}

Through the statistical Wigner-Moyal formalism described above and in \cite{moyal1949quantum}, the weak value $\tilde{p}$
takes on the role of a statistical average. However, an average is
scientifically worthless unless one can provide an associated spread.
It is well established that the variance, i.e. the second central moment of
a distribution, is equal to the second cumulant of that distribution.
With some effort, the same can be shown of conditional variances and
conditional cumulants \cite{Bartlett}, and extended to those of pseudodistributions
like the Wigner function \cite{moyal1949quantum}.

\subsubsection{Derivation\label{derivation}}

Using the characteristic function associated
to $\mathcal{W}\left(x,p\right)$ and the polar decomposition $\psi=\sqrt{\rho}e^{iS}$,
the conditional cumulants of this distribution
can be written for even cumulants ($n=2,4,\ldots$) as \cite{moyal1949quantum}

\begin{equation}
\kappa_{n}|x=\left(\frac{\hbar}{2i}\right)^{n}\left(\frac{\partial}{\partial x}\right)^{n}\ln\rho\left(x\right),
\label{eq:evencumulants}
\end{equation}
and for odd cumulants ($n=1,3,\ldots$) as
\begin{equation}
\kappa_{n}|x=\left(\frac{\hbar}{2i}\right)^{n-1}\left(\frac{\partial}{\partial x}\right)^{n}\hbar S\left(x\right).
\label{eq:oddcumulants}
\end{equation}
The derivation of this result has been reproduced in full as an Appendix.

By definition, the first two conditional cumulants $n=1,2$ are also
the first two conditional moments,
and one finds a simple relation \cite{moyal1949quantum},
\[
\tilde{p}=\frac{\partial\left(\hbar S\right)}{\partial x},
\]
for the ($n=1$) average $\tilde{p}=\mathbb{E}_{p}\left(p|x\right)$. Simple comparison of this expression with \eqref{eq:weakvaluep} should confirm that $\tilde{p} = \mathrm{Re}\left(\inlineweak{\hat{p}}\right)$. The $n=2$ expression then defines the weak variance of the random variable $p|x$:
\begin{equation}
\mathbb{V}_{p}\left(p|x\right)=-\frac{\hbar^{2}}{4}\left(\frac{\partial}{\partial x}\right)^{2}\ln\rho\left(x\right).
\label{eq:def_by_minlaplnrho}
\end{equation}
Dimensional analysis confirms that this quantity has units of $\left[\mathrm{momentum}\right]^{2}$,
as expected of a (conditional) momentum variance. Because of the clean splitting between phase-dependent odd cumulants and amplitude-dependent even cumulants, the weak variance does not depend on the wavefunction phase, except indirectly (via the time evolution of the amplitude).

\subsubsection{Other expressions for the Weak Variance}

This second conditional cumulant is also expressible directly as the second central conditional moment of the conditional distribution $\mathcal{W}\left(p|x\right)$: 
\begin{equation}
\mathbb{V}_{p}\left(p|x\right)=\int\mathcal{W}\left(p|x\right)\left(p|x-\mathbb{E}_{p}\left(p|x\right)\right)^{2}dp.
\label{eq:wigner_weak_variance}
\end{equation}

Furthermore, as we shall demonstrate in \textsection \ref{twoonetwo}, the weak variance also takes the following form, more readily related to weak values:
\[
\weakvar = \frac{1}{2}\mathrm{Re}\left(\weak{\hat{p}^2} -\left(\weak{\hat{p}}\right)^2\right).
\]
This last expression shows that despite our use of unphysical conditional random variables $p|x$ in its derivation, the weak variance is indeed related to the intuitive form $\langle p^2\rangle - \langle p\rangle^2$ of a variance. It also shows that the weak variance is \emph{measurable} in an experimental setting.

\subsection{The Uncertainty Principle}

It is possible to relate the conditional variance $\weakvar$ to the unconditional variance $\mathbb{V}\left(p\right)$ that appears in the uncertainty relations.

\subsubsection{The Law of Total Variance}

One may readily verify that the law of total expectation,
\begin{equation}
\mathbb{E}\left(p\right)=\mathbb{E}_{x}\left(\mathbb{E}_{p}\left(p|x\right)\right)=\mathbb{E}_{x}\left(\tilde{p}\right),
\end{equation}
reads in our quantum setting
\begin{equation}
\left\langle \psi\right|\hat{p}\left|\psi\right\rangle =\int\left|\psi\right|^{2}\tilde{p}dx.
\label{eq:iteratedexpectation}
\end{equation}

In the same way that one derives the law of total expectation, one may derive the Law of Total Variance:
\begin{eqnarray}
\mathbb{V}\left(p\right) & = & \mathbb{E}_{x}\left(\mathbb{V}_{p}\left(p|x\right)\right)+\mathbb{V}_{x}\left(\mathbb{E}_{p}\left(p|x\right)\right) \\
\nonumber & = & \mathbb{E}_{x}\left(\weakvar\right)+\mathbb{V}_{x}\left(\tilde{p}\right),
\label{eq:Total_Variance}
\end{eqnarray}
where $\mathbb{V}_{x}\left(\tilde{p}\right)$ is the variance of the weak
value $\tilde{p}$ as a function of $x$, and where $\mathbb{E}_{x}\left(\mathbb{V}\left(p|x\right)\right)$
is the average value of the weak variance (as a function of $x$).
$\mathbb{V}_{x}\left(\tilde{p}\right)$ can be simply given by the marginal variance
\begin{equation}
\mathbb{V}_{x}\left(\tilde{p}\right)=\int\mathcal{W}\left(x\right)
\left(\tilde{p}-\mathbb{E}_{x}\left(\tilde{p}\right)\right)^{2}dx,
\label{eq:VarWeakValue}
\end{equation}
where the law of iterated expectation \eqref{eq:iteratedexpectation}
gives $\mathbb{E}_{x}\left(\tilde{p}\right)=\left\langle \psi\right|\hat{p}\left|\psi\right\rangle $.
Hence, $\mathbb{V}_{x}\left(\tilde{p}\right)$ measures the standard
deviation of the weak value $\tilde{p}$ from the usual expectation
value $\left\langle \hat{p} \right\rangle$ as the postselection position $x$ changes. Particularly, it does not measure the variance of $p|x$ (or of any other random variable):
the variable $\tilde{p}$ in the variance $\mathbb{V}_{x}$ is deterministically defined by
\eqref{eq:VarWeakValue}. The nonweak variance $\mathbb{V}_{x}\left(\tilde{p}\right)$
contributes to the total variance of $\hat{p}$ of the uncertainty
principle, but it is not itself a manifestation of quantum randomness.

By contrast the conditional variance,
\[
\mathbb{V}_{p}\left(p|x\right)=\int\mathcal{W}\left(p|x\right)\left(p|x-\mathbb{E}_{p}\left(p|x\right)\right)^{2}dp,
\]
is truly the variance of a random variable: it says how far from
the mean $\mathbb{E}_{p}\left(p|x\right) = \tilde{p}$ one may expect to find the
random variable $p|x$. Unlike the total variance or the variance
of $\tilde{p}$, this ``scedastic function'' depends on the postselection
$x$. It is in this sense that we identify it as the weak variance.

These two parts of the total variance allow interesting limits where their contributions are negligible one with respect to the other. For example, in stationary states there is no variability in $\tilde{p}(x)$ by construction, and the total variance of the state is supported entirely by the weak variance. A more detailed discussion of these limits, and their relevance to the classical limit, is postponed until the relevant properties of the weak variance $\weakvar$ are introduced (\textsection \ref{ClassicalLimit}).

\subsubsection{Apparatus vs Fundamental Weak Variances}

This connection between the weak variance and the momentum variance reconfirms that the weak variance is central to understand the predictions of QM. Oddly, we find that the weak variance is related to the fundamental momentum variance $\mathbb{V}(p)$ \emph{of a state} (as described in the Robertson uncertainty relations) rather than the momentum change $\Delta p$ when a state is \emph{disturbed} (as originally described by Heisenberg). 

Since $\weakvar$ is given by differentiation of the wavefunction amplitude \eqref{eq:def_by_minlaplnrho}, it depends only on the values of the wavefunction amplitude in the neighbourhood of the postselection point $x$. This choice of postselection point is not apparatus-dependent, although issues such as pointer variance for the postselection variable will certainly affect the analysis of any experimental study.

The weak variance is further distinguished from an `apparatus quantity' by the fact that it is not an `observable', in the sense that it does not correspond to a Hermitean operator (cf \textsection \ref{noclassicalobservable}) or equivalently (in the presentation adopted above) to a real random variable\label{noquantumobservable}. Instead, as a cumulant of the Wigner function, it describes the state of the system, irrespective of how it is measured.

\newpage

\section{Connection to other formalisms}

One can recognise \cite{WignerHydro}
the probability flux density $\int p\mathcal{W}\left(x,p\right)\; dp=mj$
in \eqref{eq:ExpandedWignerDefinition}, and re-express this equation
using the marginal \eqref{eq:marginalWigner}, as

\begin{equation}
\frac{\tilde{p}}{m}=\frac{j\left(x,t\right)}{\left|\psi\right|^{2}}.
\label{eq:hydrodynamicVelocity}
\end{equation}
The trajectories of the Hydrodynamical and Bohmian pictures are generated
from \eqref{eq:hydrodynamicVelocity}, by setting a velocity $\dot{x}=\tilde{p}/m$
and solving the resulting ODE for $x\left(t\right)$. We will present only those elements of these formalisms that are of interest to our weak variance study; a more detailed account of this `embedding' of configuration space descriptions of QM into phase space is given in \cite{takabayasi1954formulation}.

The derivation of the real part of the momentum \begin{equation}
\tilde{p}=\partial_x\left(\hbar S\right)\label{eq:PhaseDefinition}
\end{equation} from the Schr\"odinger
Equation is present in almost every work related to the Hydrodynamical
or de Broglie-Bohm formalisms since 1927 \cite{madelung1927quantentheorie}, therefore no detail will be given here: The wavefunction is polar-decomposed, derivatives of the components split the Schr\"odinger equation into real and imaginary parts, from
which \eqref{eq:PhaseDefinition} is obtained by either appealing to an analogy
between the resulting equations and the Euler Equations (in the hydrodynamical setting) or to an analogy between these equations and the Hamilton-Jacobi Equation (in the Bohmian setting).

\subsection{Bohmian approach to the Weak Variance}

\subsubsection{Prelude}

In the quantum Hamilton-Jacobi formalism \cite{HiddenVarsBohm1,HiddenVarsBohm2,bohm1993undivided,holland1995quantum}, Hamilton's Principal Function
is $\mathcal{S}\left(x,t\right)=\hbar S\left(x,t\right)+\mathrm{const.}$,
such that the polar decomposition \eqref{eq:PhaseDefinition} becomes
the Hamilton-Jacobi definition of a canonical momentum
\begin{equation}
\tilde{p}=\partial_{x}\mathcal{S}.
\label{eq:Action Definition}
\end{equation}
Since $\tilde{p}$ is not the classical canonical momentum, but an average, the theory
remains fully quantum mechanical \cite{WeakNotBohm}. Particularly, the Hamiltonian
\begin{equation}
\mathcal{H}_\mathrm{Bohm} = -\frac{\partial \mathcal{S}}{\partial t} = \frac{\tilde{p}^{2}}{2m}+V+Q\label{eq:defBohmHamiltonian}
\end{equation}
has a quantum potential $Q$ \cite{HiddenVarsBohm1,HiddenVarsBohm2}: 
\begin{equation}
Q=-\frac{\hbar^{2}}{2m}\frac{\partial_x^{2}\left|\psi\right|}{\left|\psi\right|}.
\label{eq:Quantum Potential}
\end{equation}

The form \eqref{eq:def_by_minlaplnrho}
can be re-expressed (using $R=\left|\psi\right|=\sqrt{\rho}$) as:
\begin{equation}
\mathbb{V}_{p}\left(p|x\right) = -\frac{\hbar^2}{4} \frac{\partial}{\partial x} \left( \frac{\partial_x \rho}{\rho} \right)= -\frac{\hbar^{2}}{2} \frac{\partial}{\partial x} \left(\frac{\partial_x R}{R}\right).
\label{eq:div_of_imaginary}
\end{equation}

The imaginary part of the weak value \eqref{eq:weakvaluep} is
\begin{equation}
\mathrm{Im}\left[_{x}\hat{p}_{\psi}\right]=-\frac{\hbar}{2}\frac{\partial_x \rho}{\rho}=-\hbar\frac{\partial_x R}{R},
\label{eq:imaginary_part}
\end{equation} where the experimental $\eta$ term from \eqref{eq:experimental_eta} is suppressed. Equation \eqref{eq:div_of_imaginary} then indicates that the weak variance is proportional (with a dimensionful factor of $\hbar/2$) to the divergence of this imaginary part.

Applying the divergence operator in \eqref{eq:div_of_imaginary}
gives
\begin{equation}
\mathbb{V}_{p}\left(p|x\right)=-\frac{\hbar^2}{4} \left( \frac{\partial_x^2 \rho}{\rho} - \frac{(\partial_x \rho)^2 }{\rho^2} \right) =-\frac{\hbar^{2}}{2} \left(\frac{\partial_x^{2} R}{R} - \left(\frac{\partial_x R}{R}\right)^{2}\right),
\label{eq:curvature}
\end{equation}
and by recognising each term above (with \eqref{eq:Quantum Potential} and \eqref{eq:imaginary_part}),
we find 
\begin{equation}
\mathbb{V}_{p}\left(p|x\right)=\frac{\left[\mathrm{Im}_{x}\hat{p}_{\psi}\right]^{2}}{2} + mQ. \label{eq:cumulant=00003DQ+kinetic}
\end{equation}

This nontrivially relates the weak variance to the quantum potential of the Hamilton-Jacobi picture, providing further evidence of the physical importance of $\mathbb{V}_{p}$. The imaginary part of the weak value also appears in the expression above. By combining \eqref{eq:div_of_imaginary}, \eqref{eq:imaginary_part}, and \eqref{eq:cumulant=00003DQ+kinetic} we obtain a nonlinear differential (Riccati) equation fixing the form of $\mathrm{Im}_{x}\hat{p}_{\psi}$ at any given time,\begin{equation}
\frac{d\;\mbox{\ensuremath{\mathrm{Im}{}_{x}\hat{p}_{\psi}}}}{dx} - \frac{1}{\hbar}\left[\;\mathrm{Im}_{x}\hat{p}_{\psi}\right]^{2} = \frac{2m}{\hbar}Q(x) \label{eq:Riccati}
\end{equation}
This will be an important element in our discussion of the classical limit (\textsection \ref{ClassicalLimit}).

\subsubsection{An expression for $\weakvar$ in terms of weak values \label{twoonetwo}}

Consider the weak value of momentum squared. Substituting once again the polar form $\psi = |\psi|e^{iS}$ and $\hat{p}^2 = -\hbar^2 \partial_x^2$ gives

\begin{equation}
 \mathrm{Re}\left(\weak{\hat{p}^2}\right) = -\hbar^2 \mathrm{Re} \left(\frac{ \partial_x^2 \psi}{\psi} \right)= ( \hbar \partial_x S)^2 + 2mQ. \label{eq:weaknoncentralsecondmoment}
\end{equation}
Recognising $\hbar \partial_x S$ as the real part of the weak value of momentum, we can re-express the quantum potential $Q$ in a well-known variance-like form (up to a scaling by $2m$):

\begin{equation}
Q = \frac{1}{2m} \left( \mathrm{Re}\left(\weak{\hat{p}^2}\right) - \left[\mathrm{Re}\left(\weak{\hat{p}}\right)\right]^2 \right).
\label{eq:QAsWeakValue}
\end{equation}

Now consider the weak value of momentum, squared:
\[ \mathrm{Re}\left(\left(\weak{\hat{p}}\right)^2\right) = \mathrm{Re}(\inlineweak{\hat{p}})^2 - \mathrm{Im} (\inlineweak{\hat{p}})^2 = ( \hbar \partial_x S)^2 - \mathrm{Im} (\inlineweak{\hat{p}})^2.\]

Clearly, we can subtract this from \eqref{eq:weaknoncentralsecondmoment} above to obtain  \begin{eqnarray}
2\left(mQ + \frac{\mathrm{Im} (\inlineweak{\hat{p}})^2}{2}\right) &=& \mathrm{Re}\left(\weak{\hat{p}^2}\right) - \mathrm{Re}\left(\left(\weak{\hat{p}}\right)^2\right) \\
\weakvar &=& \frac{1}{2}\mathrm{Re}\left(\weak{\hat{p}^2} -\left(\weak{\hat{p}}\right)^2\right),
\label{eq:WeakVarAsWeakValue}
\end{eqnarray} by use of \eqref{eq:cumulant=00003DQ+kinetic} and linearity of $\mathrm{Re\left(\cdots\right)}$. This expression depends on both the first order ($\inlineweak{\hat{p}}$) and second order ($\inlineweak{{\hat{p}^2}}$) weak values, and resolves the ordering ambiguity, discussed in the Introduction, between taking the real part and taking the square of the first order weak value (compare \eqref{eq:QAsWeakValue} and \eqref{eq:WeakVarAsWeakValue}).

\subsubsection{The weak variance is not a weak value}

We have reached an expression for $\weakvar$ which superficially has the form $\langle p^2 \rangle - \langle p\rangle^2$. Let it be noted, however, that this is not a construction in terms of moments of a distribution: the complex weak value $\inlineweak{\hat{p}}$ is not a moment, it is the real part $\tilde{p} = \mathrm{Re}(\inlineweak{\hat{p}})$ that plays this role.

Consider, now, expanding the square in \eqref{eq:wigner_weak_variance} to obtain\[
\weakvar = \left( \int p^2 \mathcal{W}(p|x) dp \right) -  \mathbb{E}(p)^2 = \mathbb{E}_p(p^2|x) - \mathbb{E}(p)^2.
\] This also almost looks like $\langle p^2 \rangle - \langle p\rangle^2$, however the two expectations belong to different (pseudo)distributions ($\mathcal{W}(p|x)$ and $\mathcal{W}(x,p)$ respectively). Using the law of total expectation, we could write both terms using $\mathbb{E}_p$:\[
\weakvar = \mathbb{E}_p(p^2|x) - \mathbb{E}_p(\mathbb{E}_y(y|p))^2,
\] where $y|p$ is (formally) a position postselected on momentum. This averaging over all positions $y$ shows that despite being locally valued (in terms of the postselection $x$), $\weakvar$ encompasses information from the entire configuration space. This is not surprising if one remembers that the wavefunction solves a Schr\"odinger equation over this same space.

Quantum mechanical expectations are calculated in the Wigner formalism as integrals over the entire phase space, \begin{equation}
\langle \hat{A} \rangle = \int dx \int dp A(x,p) \mathcal{W}(x,p),
\end{equation} where $A(x,p)$ is the classical observable and Weyl operator ordering is assumed \cite{PhaseSpaceOverview}. One may then define configuration space densities
\begin{equation}
A_\mathrm{cs} = \int dp A(x,p) \mathcal{W}(x,p),
\end{equation} which give the expectation $\langle \hat{A} \rangle$ by direct integration, and `local densities' \cite{WignerHydro,kaniadakis2002statistical,CookBook}
\begin{equation}
A_\mathrm{loc} = |\psi|^{-2} A_\mathrm{cs} = \int dp A(x,p) \mathcal{W}(p|x) = \mathrm{Re}\left(\weak{\hat{A}}\right),
\end{equation} which give the expectation by weighting the integration by the probability distribution $|\psi|^{2}$. By writing our classical observable as a polynomial $A(x,p)=\sum_n f_n(x) p^n$, we find that these local densities are nothing more than linear combinations of conditional moments / weak values of $p^n$. We have already seen \eqref{eq:WeakVarAsWeakValue} that the weak variance is quadratic in the weak value $\inlineweak{\hat{p}}$, therefore it is not a local density and no classical observable associated to it exists\label{noclassicalobservable}. Furthermore this suggests (since classical observables are quantum observables too) that there is no weak variance operator $\widehat{\mathbb{V}_p}$, as argued on interpretational grounds in \textsection \ref{noquantumobservable}. Therefore, the weak variance is not a weak value.

\subsection{Hydrodynamical approach to the Weak Variance}

The real part $\tilde{p}$ of the weak value is the (Euler-picture) momentum of the Madelung fluid at position $x$. Classically, the fluid momentum is the (local) average momentum of a particle in the fluid; quantum mechanically, $\tilde{p}$ retains this qualitative feature
of being a local average as discussed above.

In the Hydrodynamical picture, one can derive \cite{WignerHydro,takabayasi1954formulation} a pressure tensor using the same logic \cite{madelung1927quantentheorie} from which one derives the fluid momentum \eqref{eq:PhaseDefinition}. We obtain in one dimension (cf. \cite{WignerHydro}, Appendix C):\begin{equation}
P = \frac{1}{m}\int (p-\tilde{p})^2 \mathcal{W}(x,p) \mathrm{d}p= \frac{\rho}{m}\left(-\frac{\hbar^2}{4} \partial_x ^2\ln(\rho) \right).
\label{eq:isotropicpressure}
\end{equation} Combining the pressure above and the equation of state of the Madelung fluid $P=\rho k_B T$ gives a definition for the thermal energy \cite{WignerHydro}:
\begin{equation}
k_{B} T=-\frac{\hbar^{2}}{4m}\partial_x^{2}\ln\rho = \frac{\weakvar}{m}.
\label{eq:WeakTemperature}
\end{equation}

There is an intuitive picture relating fluid temperatures to momentum
variances: In the kinetic theory of gases, a higher temperature corresponds
to a wider Maxwell-Boltzmann distribution of momenta, according to
$\sigma^{2}\propto mk_{B}T$. The corresponding quantum relation \eqref{eq:WeakTemperature}
preserves this intuition at the fluid level, for a temperature distribution
$T=T\left(x\right)$ over postselection $x$.

This expression \eqref{eq:isotropicpressure} should be compared with the \emph{independently motivated} expression \eqref{eq:def_by_minlaplnrho} for the weak variance. In this light, the equation of state may be justified (rather than postulated) by writing the Wigner function from \eqref{eq:isotropicpressure} in terms of its marginal, using \eqref{eq:def_by_minlaplnrho} to find the solution $P=\frac{\rho}{m}\weakvar$, and then invoking this analogy to kinetic theory to link $P$, $\rho$ and $T$ by an equation of state.

Physical potentials tend to arise from interactions (mediated by gauge fields in the Standard Model or the metric field in General Relativity), but the quantum `potential' $Q$ puzzlingly appears to pop out of the Schr\"odinger equation. Interpreting the weak variance as a thermal quantity pushes the `interactions' which give rise to $Q$ into a hypothetical microstate description from which postselected quantum mechanics emerges thermodynamically, potentially resolving this philosophical worry.

We purposefully do not propose at this time any underlying statistical thermodynamical picture, instead pointing the reader to the discussion of "subquantum theory" in \cite{WignerHydro}, which suggests that one should analyse the properties of the Wigner Function (such as, in this study, its weak variance) in order to gather some information about such a theory (cf. \textsection \ref{negativetemperatures}). It is a strength of thermodynamics that it does not formally require any model for microstates.

\subsection{Measurement of the Weak Variance}

We have shown that the weak variance can be described without reference to the properties of a specific apparatus, using only the description of the system's state through the Wigner quasidistribution or the wavefunction $\psi=Re^{iS}$. However, combining the many equivalent expressions developed in this section with the $\eta$ parameterisation of the imaginary part \eqref{eq:experimental_eta}, shows that this quantity should in fact be easy to measure in a number of experimental settings. \label{manymeasurements}

Unsurprisingly, the weak variance of momentum can be measured using weak measurements of momentum. The form \eqref{eq:WeakVarAsWeakValue} is available if one has access to the second-order weak value $\inlineweak{\hat{p}^2}$. With a good calibration of the apparatus' $\eta$ and a good resolution over $x$, \eqref{eq:div_of_imaginary} can also be used.

More interestingly, weak measurements of momentum are not necessary to quantify the weak momentum variance. Given the probability density $\rho\left(x\right)$ determined from a series of position measurements, one may simply use the $-\partial_x^{2}\ln\left(\rho\right)$ form to find the weak variance at any position, without performing a single momentum measurement, weak or otherwise. A combination of weak momentum measurements and projective position measurements (possibly the same measurements used for postselection) can be used to cross-check the previous two measurements using a single dataset.

One might also verify experimentally the law of total variance
\eqref{eq:Total_Variance}: for example, one can measure the total variance $\mathbb{V}\left(p\right)$ from projective momentum measurements, and one can measure each of the partial variances from weak momentum measurements.

Having these many experimentally independent ways to measure $\mathbb{V}_{p}\left(p|x\right)$ should allow this formalism to be stringently tested by contemporary instruments (modulo a calibration or model for the apparatus parameter $\eta$).

\newpage

\section{Negative Weak Variances}

Quantum Mechanics predicts that in some regions, $\mathbb{V}_{p}\left(p|x\right)<0$;
this odd behaviour is known to the literature, but often glossed over
(cf. e.g. the footnote to the Appendices of our primary source \cite{moyal1949quantum}).
The Bohmian and Hydrodynamical formalisms even allow the `weak
trajectories' $x\left(t\right)$ to cross from regions where $\mathbb{V}_{p}\left(p|x\right)$
is positive to regions where it is negative. It is not sufficient to say `weak values can lie outside of the eigenspectrum' to explain why a variance can turn out to be negative.

Furthermore, this is not just an odd but unmeasurable feature of the theory, like
the negative probabilities of realisations of $p|x$ \cite{moyal1949quantum,muckenheim1986review,groenewold}.
The second cumulant may, as we have just argued, be \emph{measurably} negative.

\subsection{Explaining $\mathbb{V}_{p}\left(p|x\right)<0$}

The reason we identify $\weakvar$
as a (conditional) variance is that it appears as the second term
in the expansion of the (conditional) characteristic function into cumulants \eqref{eq:evencumulants}. We therefore consider to what extent such an interpretation for this quantity is warranted beyond the (experimentally successful \cite{ObservedBohmPaths}) identification of the first cumulant as a conditional average.

The conditional variance is given by the second central conditional
moment of $\mathcal{W}$ \eqref{eq:wigner_weak_variance}: 
\begin{equation}
\weakvar=\int\mathcal{W}\left(p|x\right)\left(p-\mathbb{E}_{p}\left(p|x\right)\right)^{2}dp, \label{eq:wigner_weak_variance_repeat}
\end{equation}
which may be contrasted with the marginal moment \eqref{eq:VarWeakValue}.
The term $\left(p|x-\tilde{p}\right)^{2}$ is the square of a real quantity,
so it must be positive. Any negativity in $\weakvar$
must therefore come from the distribution $\mathcal{W}\left(p|x\right)$; and
indeed, the conditional Wigner function can take negative values \cite{moyal1949quantum}. This should be contrasted with $\mathbb{V}_{x}\left(\tilde{p}\right)$,
which depends on the marginal distribution $\mathcal{W}(x)=\left|\psi\right|^{2}\geq0$
and thus remains positive.

\subsection{Wigner Negativity and Standard Deviations \label{nostddev}}

Even though the conditional variance may be negative, allowing imaginary
standard deviations of $p|x$ would violate the requirement that $p,x\in\mathbb{R}$ (i.e. the requirement that realisations of a real random variable cannot be complex). A standard deviation defined using an \emph{ad hoc} extension to negative values,
\begin{equation}
\sigma_{p|x}=\sqrt{\mathrm{abs}\left(\mathbb{V}_{p}\left(p|x\right)\right)},
\label{eq:stddev1}
\end{equation}
satisfies $\sigma_{p|x}\in\mathbb{R}^{+}$, such that all our random variables are real, as required. This extension manifestly reproduces the usual standard deviation when $\weakvar\geq0$. This last point is necessary not only for mathematical consistency, but also for consistency of the thermodynamical picture \eqref{eq:WeakTemperature}: The standard deviations of momentum in kinetic theory are obtained from the Maxwell-Boltzmann distribution $\sigma\propto\sqrt{mk_{B}T}$.

However, the conditional distribution is a pseudodistribution, invalidating the simple expression \eqref{eq:stddev1}. Consider the integral form \eqref{eq:wigner_weak_variance_repeat} of the weak variance: even when $\weakvar = 0$, there can be a spread of $p|x$ away from the weak value, since the negative and positive parts of the pseudodistribution $\mathcal{W}\left(p|x\right)$ may cancel under the integral defining a cumulative distribution, without any information given about the squared-term $(p|x - \tilde{p})^2$. Expressed more formally, deviation-bounding theorems such as the Chebyschev inequality do not apply to quasidistributions. This possibility for \emph{stochasticity at zero variance} is further emphasised by the fact that higher $n\geq2$ cumulants (\ref{eq:evencumulants},\ref{eq:oddcumulants}) are generally nonzero, indicating that the pseudoprobability distribution is generally nontrivial.

One might then attempt to push the absolute value inside the integral that defines $\weakvar$ to take care of this problem, defining instead our standard deviation with
\begin{equation}
\sigma_{p|x}^2 = \int \mathrm{abs}(\mathcal{W}\left(p|x\right))\left(p|x-\mathbb{E}_{p}\left(p|x\right)\right)^{2}dp.
\label{eq:stddev2}
\end{equation}

By virtue of the triangle inequality $\left|\int A dx\right| \leq \int \left|A\right| dx$, this standard deviation gives at least as large a spread in $p|x$ as \eqref{eq:stddev1}. Furthermore, it clearly reproduces \eqref{eq:stddev1} whenever $\mathcal{W}\left(p|x\right) \geq 0$, and so inherits all the desirable properties discussed above.

However, this approach is similarly unacceptable. The expectation $\mathbb{E}_{p}$ is also an integral over $p$, yet if we want to consider deviations away from $\tilde{p}$ then we cannot replace the conditional distribution by its absolute value in this expectation integral: changing the distribution changes the random variable of which we consider the mean! Furthermore, the correct normalisation of the quasidistribution $\int dp \mathcal{W}\left(p|x\right) = 1$ guarantees that $\mathrm{abs}(\mathcal{W}\left(p|x\right))$ is not normalised, and so it is not a proper probability distribution either.

In light of these failures, it is even possible to dispute whether a standard deviation is a sensible quantity for quasidistributions. Unlike the variance, which pertains to the distribution, the standard deviation is a deviation of \emph{realisations} of a random variable from their mean. As discussed previously, realisations of $p|x$ can occur with negative probability: building a sample standard deviation from these phantasmagoric realisations would be meaningless as the final step of a calculation, even in a quasistatistical outlook.

\subsection{Shape of $\psi$}

In the description of Bose-Einstein superfluids, a quantum-potential-like term appears in the polar decomposition of the Gross-Pitayevski equation. In this context, $Q$ is interpreted as the kinetic energy cost of curvature of the condensate wavefunction -- it costs energy to squeeze a wavepacket.

Given that the weak variance depends in part on the quantum potential \eqref{eq:cumulant=00003DQ+kinetic}, one might wonder whether (by analogy) $\weakvar$ is related to the curvature of the wavefunction.

We now consider how the curvature of the wavefunction amplitude, $R=|\psi|$, determines the sign of the weak variance. The relevant quantity is the convexity $\partial_x^{2}R$ of the amplitude, which is positive for convex intervals of $|\psi|$ and negative for concave intervals.

\subsubsection{Non-Nodal Points}
We use the decomposition \eqref{eq:curvature} to solve the inequality
$\mathbb{V}_{p}\left(p|x\right)\geq0$. We note that $R$ is nonnegative by
definition: for non-nodal points (i.e. points $x$ such that $R(x)\neq 0$) the weak variance is nonnegative whenever the
convexity satisfies
\begin{equation}
\frac{\left(\partial_x R\right)^{2}}{R}\geq\partial_x^{2}R.
\label{eq:small_convexity}
\end{equation}

The left-hand side of \eqref{eq:small_convexity} is positive, so concave regions ($0>\partial_x^{2}R$) always have positive weak variances (independently of the system under consideration). Regions that have a very
small concavity are also very likely to satisfy \eqref{eq:small_convexity}.

Maxima of the wavefunction amplitude are concave so $\weakvar>0$. Near local minima of $R(x)$, the left-hand side of \eqref{eq:small_convexity} vanishes and the inequality is never satisfied (since $\partial_x^2 R >0$): Hence we expect $\weakvar\leq0$ near local minima (again, independently of the specific system under consideration).

In the extremal case, piecewise exponential amplitudes $R\left(x\right)=\exp(\pm kx)$ always give $\mathbb{V}_{p}\left(p|x\right) = 0$. These amplitudes are observed e.g. for particles tunnelling into a constant potential barrier.

Since the position-dependence of the amplitude $R(x)$ is easy to measure with projective position measurements alone, this inequality can serve as a test for the sign of $\weakvar$ even in an experiment not optimised to calculate the weak variance. We recommend that experimental probes into negative variances should occur near non-nodal minima of the probability density, since
the convexity in other regions may not support negative weak variances.

\subsubsection{Nodal Points}

A slight complication arises around nodal points $R\rightarrow0$ of the
wavefunction: The weak variance at the node itself is infinite, given the logarithmic singularity at $\ln(0)$ that appears in \eqref{eq:def_by_minlaplnrho}.

To show this, we note that continuous differentiability of $\psi$ and $\psi^{*}$ enforces $\partial_x R\rightarrow0$ around nodes. The expansion of the amplitude $R$ about a node at any fixed time is then
\[
R(x) = ax^2+\mathcal{O}(x^3),
\]
where the condition $\nabla^{2}R>0$ (that a node is a minimum of the amplitude) gives $a>0$. Around this node, the weak variance \eqref{eq:def_by_minlaplnrho} tends to:
\begin{equation}
-\frac{\hbar^2}{2}\partial_x^2\ln ax^2=-\hbar^2 \partial_x^2\ln x =-\hbar^2 (-\frac{1}{x^2}) \rightarrow +\infty. \label{eq:nonodes}
\end{equation} It is simple to see that this asymptotic result generalises even to pathological expansions $R=0+\cdots+0+\mathcal{O}(x^n)$. Interestingly, this behaviour for the weak variance is always strictly positive, excluding $\mathbb{V}_{p}\left(p|x\right)\leq 0$ very close to the node. This will be illustrated in the following section in a number of systems.

\subsubsection{Illustration in toy systems}
The study of coherent states in the context of negativity of the Wigner function is motivated by the result \cite{hudson1974wigner} that \emph{only} these states are nonnegative everywhere.

Consider the coherent states of the Quantum Harmonic Oscillator. We must check (for consistency) that $\weakvar>0$ (strictly) in these states.
The weak variance is 
\[
\mathbb{V}_{p}\left(p|x\right)=-\frac{\hbar^{2}}{4}\partial_x^{2}\left[\ln\left(\left(\frac{m\omega}{\hbar\pi}\right)^{\frac{1}{4}}e^{-\frac{m\omega}{2\hbar}\left(x-\left\langle \hat{x}\left(t\right)\right\rangle \right)^{2}}\right)^{2}\right],
\]
and, after many cancellations, one obtains the simple expression
\[
\mathbb{V}_{p}\left(p|x\right)=+ m \frac{\hbar\omega}{2}.
\]

This is indeed positive for all $x$, as expected, and manifestly
has the required dimensions for a momentum variance:\[
\left[\mathrm{mass}\right]\times\left[\mathrm{energy}\right]=\left[\mathrm{momentum}\right]^{2}.
\]

We note that $\mathbb{V}_{p}\left(p|x\right)/m$ is equal to the
zero-point energy $\hbar\omega/2$ in this minimum-uncertainty state.
In fact, for any energy eigenstate of the harmonic oscillator, we find a
zero-point contribution to the variance:
\begin{equation}
\frac{\mathbb{V}_{p}\left(p|x\right)}{m}=+\frac{\hbar\omega}{2}-\frac{\hbar\omega}{2}\left[\left(\frac{\partial}{\partial y}\right)^{2}\ln H_{n}\left(y\right)\right]\, ,\; y=\sqrt{\frac{m\omega}{\hbar}}x, \label{eq:eigenstate}
\end{equation}
and this contribution remains even for arbitrary superpositions
thereof.

For an energy eigenfunction of a particle in a box (respectively in a $-\alpha/\left|x\right|$ potential), we obtain
\begin{eqnarray}
\frac{\mathbb{V}_{p}\left(p|x\right)}{m} &=& + E_{n} \left[\csc^2(\frac{n\pi}{L}[x-x_{0}])
\label{eq:varianceinabox} \right] \\
(\mbox{resp.}) &=& \frac{\hbar^{2}}{2m}\frac{1}{x^{2}} + E_{n}\left[\left(\frac{\partial}{\partial y}\right)^{2} \ln L_{n-1}^{(1)}\left(y\right)\right],
\end{eqnarray} in terms of the energy levels $E_{n}, n\ge1$ of these two systems.
The eigenstate in a box or a harmonic oscillator is clearly not in a coherent state; yet we find that $\weakvar > 0$ for all $x$ for these toy models, because each $x$ lies in the vicinity of a maximum or a nodal point.

We note in passing that the weak variance becomes positively infinite at the nodal points of these systems, as expected from the discussion in the previous section. More interestingly, the weak variance tends (at certain postselections $x$) to the zero-point energy when the system is in its ground state. We purposefully do not propose at this time an interpretation of this finding.

\subsection{Negative Temperatures \label{negativetemperatures}}

If the weak variance is the temperature distribution \eqref{eq:WeakTemperature}
of the Madelung fluid, then the Madelung fluid can have negative temperatures.
Negative temperatures may be surprising, although their study is theoretically
well-established \cite{ramsey1956thermodynamics} and the concept
has successfully been applied to several experimental systems exhibiting population inversion. Some interesting prerequisites for negative temperatures are that the internal energy of the system must be bounded from above as well as below, and that $T=0$ requires the energy to be extremised to one of these bounds.
The ideal gas law $P=\rho k_B T$ shows that the pressure also becomes negative, precisely when the temperature is negative.

One further consequence of weak variances on the thermodynamics of the hydrodynamical formalism is interpretational. Recall that the hydrodynamic momentum $\tilde{p}=\mathbb{E}(p|x)$ is interpreted \cite{WignerHydro} as the local average momentum of the fluid particles at a position $x$. Yet, we have seen (\textsection \ref{nostddev}) that negative weak variances do not provide a standard deviation or a Chebyschev inequality for the random variable $p|x$. In the hydrodynamic language, the peculiar momentum of a Madelung fluid particle is not required to be zero, even at zero temperature! Also note that the temperature of a nodal point in the fluid (where we expect to find no fluid particles) is $+\infty$.

These are the sort of features any ``subquantum theory'' \cite{WignerHydro} should exhibit, if such a theory admits the Madelung hydrodynamics as its thermodynamic limit.

\section{Nondeterminism in Quantum Mechanics}

We have already discussed (\textsection \ref{nostddev}) how a null weak variance still allows stochasticity and nonzero standard deviations. Although $\weakvar=0$ does not imply determinism, one might nonetheless consider whether a relation between the weak variance and determinism exists. For example, the highly nonlinear form of the weak variance makes it unlikely that a null weak variance would emerge for superposition states (or, a fortiori, for entangled states). These conditions correspond exactly to the nonclassical regime, where we expect nondeterminism.

\subsection{The Classical Limit \label{ClassicalLimit}}

\subsubsection{Consistency of the Classical Limit}

It should not be surprising to find that these worrisome negative variances
disappear in the classical limit. Heuristically considering the classical limit as the limit of infinitely thin coherent states similarly guarantees $\weakvar \geq 0$ since the conditional distribution is nonnegative \cite{hudson1974wigner}.

However, a stronger statement is also true: weak variances should disappear completely in the classical limit, since classical observables are well-defined and motion is deterministic. We have even shown (\textsection \ref{noclassicalobservable}) that there is no classical observable associated to it. The quantum potential behaves like $Q\rightarrow0$ in the classical limit \footnote{One must strictly take the limit of the quantum potential for the entangled state of the system and the measurement apparatus \cite{bohm1993undivided}, but $Q\rightarrow0$ for the system alone is not seen as problematic in the literature.}. When $Q=0$, the Riccati equation \eqref{eq:Riccati} admits a trivial solution $\mathrm{Im}_{x}\hat{p}_{\psi} = 0$ since the weak momentum ($\in \mathbb{C}$) tends to the classical momentum ($\in \mathbb{R}$). Hence, using \eqref{eq:cumulant=00003DQ+kinetic}, $\mathbb{V}_{p}\left(p|x\right)\rightarrow0$ classically.

\subsubsection{Deriving the Classical Limit}

Another approach to the classical limit would be to minimise the contribution of the scedastic function $\weakvar$ in the total variance \eqref{eq:Total_Variance} by our choice of wavefunction, in order to mimic determinism. Since the weak variance does not depend on the wavefunction phase, only the probability density $\rho = |\psi|^2$ is relevant. We can express this extremisation problem in the language of calculus of variations as\begin{eqnarray}
\delta J[\rho] &=& \delta \left(\mathbb{E}_{x}\left(\mathbb{V}_{p}\left(p|x\right)\right)\right) \\
\nonumber &=& \delta \int dx \rho(x) \; \partial_x^2 \log(\rho(x)) \\
\nonumber &=& \delta \int dx \left[ \rho^{\prime \prime} - \frac{(\rho^\prime)^2}{\rho}\right] \\
\nonumber &=& \int dx \; \delta \mathcal{L}(\rho,\rho^\prime,\rho^{\prime \prime}) \\
\nonumber &=& 0,
\end{eqnarray} where unimportant constant factors are suppressed. The resulting Euler-Lagrange equation depends (albeit trivially given our specific $\mathcal{L}$) on the second derivative $\rho^{\prime \prime} = \partial_x^2 \rho$ of the probability density:\begin{equation}
\frac{\partial \mathcal{L}}{\partial \rho} - \frac{\partial}{\partial x} \frac{\partial \mathcal{L}}{\partial \rho^\prime} + \frac{\partial^2}{\partial x^2} \frac{\partial \mathcal{L}}{\partial \rho^{\prime \prime}} = \frac{\rho^{\prime \prime}}{\rho} - \frac{1}{2} \left( \frac{\rho^\prime}{\rho}\right)^2= 0.
\label{eq:EulerLagrangeExtremum}
\end{equation}
We notice that the terms of \eqref{eq:curvature} are similar to those of the Euler-Lagrange equation \eqref{eq:EulerLagrangeExtremum}. Consequently we find that in this limit, a term of our choice cancels out of \eqref{eq:curvature} to give \[
\weakvar =\left(\mathrm{Im}\left[_{x}\hat{p}_{\psi}\right]\right)^2 = -mQ.
\] However, reinserting this into \eqref{eq:cumulant=00003DQ+kinetic} results in $\weakvar=+mQ$; this is uniquely possible for $\weakvar=Q=\mathrm{Im}\left[_{x}\hat{p}_{\psi}\right]=0$. We have therefore derived that the minimum average contribution of the weak variance to the total variance is zero: negative weak variances do not allow the nonweak variance $\mathbb{V}_x(\tilde{p})$ to exceed the total variance in \eqref{eq:Total_Variance}.

In summary, by minimising the stochastic part of the total variance we can \emph{derive} the classical limit $Q\rightarrow0$ of the Hamilton-Jacobi formalism.

\subsection{$\weakvar \rightarrow 0$ as a Semiclassical Limit}

Equation \eqref{eq:div_of_imaginary} states that $\weakvar$ is the divergence of the imaginary part of the weak value: when this variance is zero, $\mathrm{Im}(\inlineweak{\hat{p}})$ is a function of time only. In this case, either \eqref{eq:cumulant=00003DQ+kinetic} or the Ricatti equation \eqref{eq:Riccati}, gives that $Q(x,t)=Q(t\,\mathrm{only}) < 0$. Hence the quantum force $\partial Q /\partial x = 0$: this is \cite{holland1995quantum} the condition that Bohmian trajectories coincide with their classical counterparts. $\weakvar \rightarrow 0$ therefore yields classical particle dynamics (on average), occurring at nonclassical energies. Of course, these nonclassical dynamics are localised to specific postselections $x$: if $\weakvar=0,\; \forall x$ then the average weak variance is also zero and we recover the classical limit as derived above.

Consider a particle tunnelling through a constant barrier: The exponential wavefunction amplitude guarantees that $\weakvar = 0$ inside the barrier, therefore the tunnelling particle lives in this semiclassical limit and will (on average) follow the path of a classical particle with a higher energy: it will go over the barrier. The fact that tunnelling particles behave semiclassically is indeed why tunnelling problems are successfully visualised and formulated in terms of particles.

\subsection{Weak variances do not `measure' a deterministic particle ontology}

It is sometimes claimed that the weak value $\mathrm{Re}(\inlineweak{\hat{p}})$ `measures' the Bohmian momentum \cite{wiseman2007grounding,ObservedBohmPaths}. A naive Bohmian might expect that the weak variance would vanish, being the variance of the nonstochastic Bohmian momenta. In this section we argue against one possible misinterpretation of these quantities: weak values (and their associated weak variances) do not measure the properties of a Bohmian particle.

The \emph{only} hidden variable of a Bohmian particle is position -- it does not even have an energy or even a momentum independently of what the mathematical formalism predicts we should measure \cite{bohm1993undivided,durr2009bohmian}. This is of course a reflection of the fact that nonposition measurements in Bohmian mechanics are contextual \cite{bohm1993undivided}. As such, a measurement of the weak value, although equal to the Bohmian momentum $\partial_x \mathcal{S}$, is not a measurement of the ``particle's momentum''; similarly, the weak variance does not measure the standard deviation of Bohmian particles' momentum (both on interpretational and quasistatistical grounds).

With the above caveat in mind, the weak variance allows a deterministic particle interpretation of the `semiclassical' limit $\weakvar \rightarrow 0$: such a (localised) Bohmian interpretation has already been shown to apply at postselection points $x$ where it is at its strongest (particles tunnelling and in the classical limit), and is guaranteed to not apply where the global Bohmian interpretation is known to break down \cite{bohm1993undivided,holland1995quantum,GlobalBohmian,SimpleGlobalBohmian} (since $\weakvar\rightarrow+\infty\neq0$ at nodal points). We emphasise that a semiclassically emergent Bohmian ontology is neither supported nor precluded by the absence of a sensible notion of standard deviation for quasidistributions, nor can it be of any instrumental relevance.

\section*{Conclusions}
The weak variance can be defined naturally and uniquely in terms of the Wigner function, singled out by insisting on a (quasi)statistical interpretation of quantum phase space. A weak variance does not define the standard deviation of individual realisations of $p|x$, which may occur with negative probability, but instead describes the system's state.

The weak variance $\weakvar$ contributes to the momentum variance of the uncertainty principle, justifies the `ideal gas' equation of state of the hydrodynamical formalism, and appears to be related to the ground state energy of a few analytically soluble systems. An expression $\mathrm{Re}(\inlineweak{\hat{p}^2}-\inlineweak{\hat{p}}^2)$ in terms of weak values was derived, which we used to argue that the weak variance is not itself a weak value.

We show that the sign of the weak variance is predetermined for local extrema of the probability density: $\weakvar$ is positive for nodal points and maxima, and negative for non-nodal extrema. We also derive some features that any hypothetical "subquantum theory" (from which quantum mechanics emerges as a thermodynamic limit) must satisfy.

We show that the weak variance is not only consistent with the classical limit, but can be used to derive the classical limit from a variational principle where the expectation of the weak variance controls the emergence of determinism. This construction also shows that the variability $\mathbb{V}_x(\tilde{p})$ of the weak value cannot exceed the total variance of the state. A semiclassical limit is derived and discussed.

Most importantly, we note that $\weakvar$ should be measurable by a number of experimental setups (\textsection \ref{manymeasurements}).

\newpage
\bibliographystyle{spphys}
\bibliography{Bib}

\newpage

\section*{Appendix: Derivation}
The characteristic function of a distribution is its Fourier transform. For the Wigner function we can write the conditional characteristic function \cite{moyal1949quantum,Bartlett} as follows:
\begin{equation}
M(\tau|x) = \int dp \mathcal{W}(p|x) e^{i\tau p / \hbar} = \frac{ \psi^*(x-\frac{\hbar}{2} \tau) \psi(x+\frac{\hbar}{2} \tau)}{\psi^*\psi(x)}.
\end{equation}
The cumulant-generating function is the logarithm of the characteristic function. This gives
\begin{eqnarray}
\ln(M(\tau|x)) &=& \ln(\frac{ \psi^*(x^-) \psi(x^+)}{\rho(x)}) = \ln(\psi^*(x^-)) + \ln(\psi(x^+)) - \ln(\rho(x)) \\
\nonumber &=& \frac{1}{2}\left[\ln(\rho(x^-)) + \ln(\rho(x^+))\right]+ i\left[S(x^+) - S(x^-)\right]  - \ln(\rho(x)),  \\
\end{eqnarray} which we can then Taylor expand to obtain the cumulants \cite{moyal1949quantum}.

By inspecting \eqref{eq:def_by_minlaplnrho}, we expect final factors of $(\hbar/2i)^n$ and an expansion in $(i\tau)^n/n!$, so we consider series of the form:\[
f(x \pm \frac{\hbar}{2i} (i\tau)) = \sum_{n=0}^\infty \left[\left(\pm \frac{\hbar}{2i}\right)^n (\partial^n f)(x) \right] \frac{(i\tau)^n}{n!},
\] where the $\left(\hbar/2i\right)^n$ come from the chain rule.

From the $i(S^+-S^-)$ term we obtain a contribution to cumulants of \[
i \left[ \left(+\frac{\hbar}{2i}\right)^n - \left(-\frac{\hbar}{2i}\right)^n \right] \partial^n S =  \begin{cases}
\left( \frac{\hbar}{2i} \right)^{n-1} \partial^n (\hbar S) &\mbox{if } n \mbox{ odd} \\
0 &\mbox{if } n \mbox{ even}
\end{cases}
\]
and from the $\ln \rho$ terms we obtain:
\[
\left[ \frac{1}{2} \left(+\frac{\hbar}{2i}\right)^n + \frac{1}{2}\left(-\frac{\hbar}{2i}\right)^n - \delta_{0n} \right] \partial^n \ln\rho = \begin{cases}
0 &\mbox{if } n = 0 \\
\left(\frac{\hbar}{2i}\right)^n \partial^n \ln\rho &\mbox{if } n \mbox{ even} \\
0 &\mbox{if } n \mbox{ odd}
\end{cases}
\]
\end{document}